\documentclass[twocolumn]{aastex631}

\newcommand{\Rm}{{\rm Rm}}

\usepackage{amsmath}
\usepackage{esint}
\usepackage{ragged2e}
\usepackage{hyperref}
\usepackage{physics}
\usepackage{bm}

\shorttitle{MHD oscillations from thermo-resistive instability in hot jupiters}
\shortauthors{Hardy et al.}

\begin{document}

\title{Magnetohydrodynamical torsional oscillations from thermo-resistive instability in hot jupiters}

\author[0000-0002-2599-6225]{Raphaël Hardy}
\affiliation{Département de Physique, Université de Montréal, Montréal, QC, H3C 3J7, Canada\\}
\affiliation{Department of Physics and Trottier Space Institute, McGill University, Montréal, QC, H3A 2T8, Canada\\}
\affiliation{Institut de Recherche sur les Exoplanètes (iREx), Université de Montréal, Montréal, QC H3C 3J7, Canada}

\author[0000-0003-1618-3924]{Paul Charbonneau}
\affiliation{Département de Physique, Université de Montréal, Montréal, QC, H3C 3J7, Canada\\}

\author[0000-0002-6335-0169]{Andrew Cumming}
\affiliation{Department of Physics and Trottier Space Institute, McGill University, Montréal, QC, H3A 2T8, Canada\\}
\affiliation{Institut de Recherche sur les Exoplanètes (iREx), Université de Montréal, Montréal, QC H3C 3J7, Canada}

\correspondingauthor{Raphaël Hardy}
\email{raphael.hardy@umontreal.ca}



\begin{abstract}

Hot jupiter atmospheres may be subject to a thermo-resistive instability where an increase in the electrical conductivity due to ohmic heating results in runaway of the atmospheric temperature. We introduce a simplified one-dimensional model of the equatorial sub-stellar region of a hot jupiter which includes the temperature-dependence and time-dependence of the electrical conductivity, as well as the dynamical back-reaction of the magnetic field on the flow. This model extends our previous one-zone model to include the radial structure of the atmosphere. Spatial gradients of electrical conductivity strongly modify the radial profile of Alfvén oscillations, leading to steepening and downwards transport of magnetic field, enhancing dissipation at depth.
We find unstable solutions that lead to self-sustained oscillations for equilibrium temperatures in the range $T_\mathrm{eq}\approx 1000$--$1200$~K, and magnetic field in the range $\approx 10$--$100$~G. For a given set of parameters, self-sustained oscillations occur in a narrow range of equilibrium temperatures which allow the magnetic Reynolds number to alternate between large and small values during an oscillation cycle. Outside of this temperature window, the system reaches a steady state in which the effect of the magnetic field can be approximated as a magnetic drag term. Our results show that thermo-resistive instability is a possible source of variability in magnetized hot jupiters at colder temperatures, and emphasize the importance of including the temperature-dependence of electrical conductivity in models of atmospheric dynamics.
\end{abstract}



\section{Introduction}\label{sec:intro}
    	Exoplanet surveys have detected nearly 500 hot jupiters (HJ), gas giants orbiting their host star with an orbit of $\lesssim 0.1 {\rm AU}$, according to the data of \href{exoplanet.eu}{exoplanet.eu} \citep{Schneider2011}. Due to their proximity to their hosts, they are exposed to extreme irradiation and are assumed to be tidally locked. Such an arrangement results in interesting atmospheric dynamics, including a dayside to nightside flow driven by the gradient of temperature \citep{Komacek2016}, an equatorial superrotation driven by Rossby and Kelvin waves \citep{Read2018,Imamura2020} resulting in the zonal displacement of the hot spot away from the substellar point, and magnetic effects due to partial ionization in the atmosphere \citep{Rogers2014b}. Indeed, HJs may well be some of the most non-axisymmetric single astrophysical bodies known to date.
    	
    	Even though the first discovery of a hot jupiter goes back a quarter of a century \citep{Mayor1995}, some features of these planets are still not fully understood. Some of these unexplained observations are the larger than expected radii \citep{Bodenheimer2001, Guillot2002, Baraffe2003, Bodenheimer2003, Laughlin2005, Thorngren2021}, differences in dayside to nightside circulation efficiency \citep{Cowan2007, Knutson2007, Crossfield2010, Cowan2012}, and the westward winds observed in some hot jupiters \citep{Armstrong2016, Dang2018,Jackson2019,Bell2019,vonEssen2020} instead of the predicted eastward direction \citep{Showman2002, Cooper2005, Showman2009, Rauscher2010, Kataria2016}, although many hot jupiters do show eastward flows \citep{Knutson2007, Knutson2009, Knutson2012, Zellem2014}.

    	A number of proposed explanations for these puzzles involve magnetism. 
    	%
        The temperatures in HJs are sufficiently high for thermal ionization \citep{Batygin2010,Perna2010a}, in particular of alkali metals, which have low first ionization energies \citep{Welbanks2019}. It has been previously shown that ionization in the atmospheres of these gas giants results in coupling of winds with magnetic field generated deeper in the planet \citep{Batygin2010,Perna2010a,Perna2010b,Menou2012a}. Magnetic interaction between the upper and lower envelopes of the HJs can slow down the atmospheric winds, and strong fields may even reverse the wind direction \citep{Rogers2014b, Rogers2017b, Hindle2019, Hindle2021a, Hindle2021b, Hardy2022, Dietrich2022}. Such magnetohydrodynamical (MHD) interaction may result in the hot spot being west of the substellar point.

        One complication in modelling magnetized HJ atmospheres is that the electrical conductivity is strongly temperature-dependent (e.g.~see Figure~2 in \citealt{Dietrich2022}). This also leads to new physics: \citet{Menou2012b} (M12) showed that the strong temperature-dependence of the electrical conductivity gives rise to a thermo-resistive instability (TRI), in particular
        in cases with weak magnetic drag, strong ohmic heating and fast winds. 
        Whereas the model presented by M12 was in a non-dynamical framework, \citet{Hardy2022} (H22) added dynamics, allowing Alfvénic oscillations to occur. They identified three different dynamical regimes of the TRI. In the first regime, the system evolves to a steady-state, either by decaying Alfvén waves or monotonically. The second regime is the burst regime, where bursts of decaying Alfvén waves happen periodically following quiescent intervals. In the third regime, Alfvén waves are continually sustained (although they found that this third regime does not occur for parameter values relevant for HJ atmospheres).

    	In this work, we further investigate the effect of the temperature-dependent electrical conductivity $\sigma$, or equivalently magnetic diffusivity (MD) ($\eta=1/\mu_0\sigma$ where $\mu_0$ is the permeability of free space). We extend the local model of H22 to include the radial structure of the HJ atmosphere. 
     Previous studies have included magnetic effects by adding an approximate treatment of magnetism to detailed atmospheric models, for example 
     either using MD profiles fixed by the background temperature \citep[][]{Rogers2014a,Rogers2014b, Rogers2017b}, or using a time-dependent MD tied to local temperature changes, but with dynamics approximated by a magnetic drag term \citep{Rauscher2013, Beltz2022}. 
     Here, we take the alternate approach of including the full dynamics and time-dependent MD but in a simplified geometry, focusing on the equatorial plane and assuming axisymmetry. Our goal is to study the coupling of dynamics and thermal evolution, and to investigate under what conditions self-sustaining oscillations (or bursts of oscillations) are excited.
     
        The paper is organized as follows. In Section \ref{sec:1dm} we describe our one-dimensional equatorial plane model, geometrically simplified but retaining both temperature-dependent MD as well as the dynamical interaction between the zonal flows and magnetic fields. Section \ref{sec:results}
        presents and discusses a selected set of simulation results, focusing on the physical parameter regimes conducive to the development of instability-driven zonal flow oscillations. We close in Section \ref{sec:discussion} by highlighting potential observable consequences. 
        

\section{One dimensional equatorial plane model}\label{sec:1dm}


    \subsection{Model setup}\label{sec:setup}

\begin{figure}[htpb]
            \centering
            \includegraphics[width=0.55\linewidth]{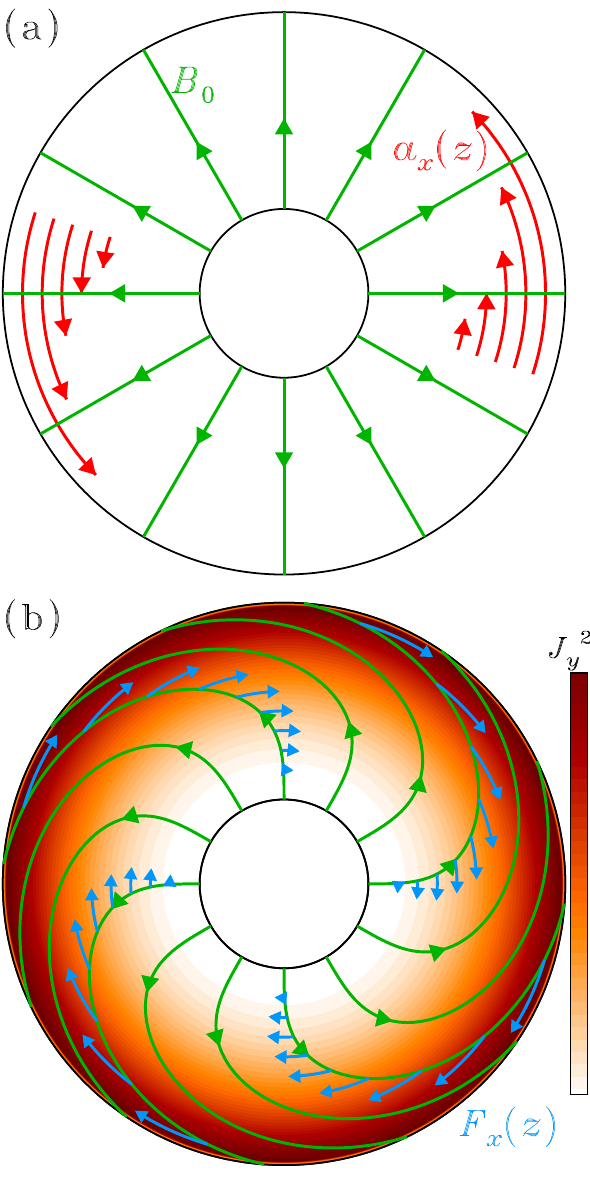}
                \caption{ {\bf (a)} Schematic representation of the assumed flow and magnetic field geometry. The aspect ratio between inner and outer cylinders has been greatly exaggerated for visual clarity. The magnetic field lines are represented in green 
                and the acceleration term 
                in red. {\bf (b)} Schematic representation of the system during its evolution. The magnetic field lines are stretched by the azimuthal flow. 
                The Lorentz force
                acting to straighten the field lines is shown in light blue. The local ohmic heating is represented by the value of $J_y^2$, where $J_y$ is the induced current, as color coded. Note that since the atmosphere is geometrically thin, we use Cartesian rather than spherical coordinates in our calculations, as described in \S~2.}
            \label{fig:1dm_cartoon}
        \end{figure}

        Working under the MHD approximation \citep{Davidson2001}, we consider a geometrically-simplified setup within the equatorial plane, depicted schematically in Figure~\ref{fig:1dm_cartoon}. Our solution domain is meant to represent the upper atmosphere of HJs, going from $1$ bar at the base to $0.01$ bar at the top. For a planet with equilibrium temperature $T_{\mathrm eq}=1000\ {\rm K}$ and surface gravity $g_p = 9.0$~m~s$^{-2}$, this corresponds to a radial extent $r_0=0.940$--$0.966~R_p$ assuming hydrostatic balance. For simplicity we assume an ideal gas of pure molecular hydrogen. As the layer is geometrically thin, we adopt the plane-parallel approximation in which cartesian coordinates $(x,y,z$) map to the zonal, latitudinal and radial directions in spherical coordinates. We solve the zonal ($x$) components of the MHD equations in the equatorial plane ($y=0$), assuming axisymmetry and that the flow is confined to the equatorial plane. We thus suppress all radial and latitudinal flow displacements, i.e.,  $u_y = u_z = 0$. 

        We express the axisymmetric magnetic field as
        \begin{equation}\label{eq:Bdef}
            {\bf B}(z)=B_x(z){\hat x}+B_0{\hat z}~,
        \end{equation}
        thus ensuring $\nabla\cdot {\bf B}=0$, and with $B_0$ setting the (fixed) strength of the radial magnetic component in the equatorial plane.         
        The assumption of a radial background magnetic field allows us to write down the simplest model possible in the equatorial plane that captures the dynamics of Alfvén oscillations (Figure~\ref{fig:1dm_cartoon}). A global aligned dipole lacks any radial component at the equator, and would require adding the latitudinal dimension to capture Alfvén oscillations. Of course, the true field geometry is likely to be more complicated than an aligned-dipole. A misaligned dipole or higher order multipoles would give radial field at the equator, and the actual magnetic field geometry could also be much more complex if a local dynamo operates in the atmospheric layer \citep{Rogers2017a,Dietrich2022}. We adopt this simplified model as a starting point to capture the physics of the interaction of the temperature-dependent MD with magnetic torques.
        
    	The reduced MHD equations solved are thus:
    	\begin{equation}\label{eq:movement}
        \frac{\partial u_x}{\partial t} = \frac{B_0}{{\mu}_0 {\bar \rho}}  \frac{\partial B_x}{\partial z}  + \frac{{\bar \mu}}{{\bar \rho}}  \frac{\partial^2 u_x}{\partial z^2}  + a_{x},
        \end{equation}
        %
        \begin{equation}\label{eq:induction}
        \frac{\partial B_x}{\partial t} = B_0  \frac{\partial u_x}{\partial z}   +  \frac{\partial \eta}{\partial z}  \frac{\partial B_x}{\partial z}  + \eta  \frac{\partial^2 B_x}{\partial z^2},
        \end{equation}
    	%
    	\begin{equation}\label{eq:temperature}
    	\begin{aligned}
        \frac{\partial T}{\partial t} &=  \frac{1}{{\bar \rho} c_p}  \frac{\partial \bar \chi}{\partial z}  \frac{\partial T}{\partial z}  + \frac{\bar \chi}{{\bar \rho} c_p}  \frac{\partial^2 T}{\partial z^2}  + \frac{1}{{\bar \rho} c_p} \frac{\partial F_{\rm irr}}{\partial z} \\& + \frac{{\bar \mu}}{{\bar \rho} c_p} \left( \frac{\partial u_x}{\partial z} \right)^2 + \frac{\eta}{{\mu}_0 {\bar \rho} c_p}  \left( \frac{\partial B_x}{\partial z} \right) ^2.
        \end{aligned}
        \end{equation}
        %
    	Equation (\ref{eq:movement}) is the $x$-component of the incompressible Navier-Stokes equation where 
        $\bar \rho$ is the density, and $\bar \mu$ is the dynamic viscosity. The fluid motions are forced by an acceleration term
        \begin{equation}\label{eq:velocity_forcing}
    	    a_{x} = \dot{v} \exp({-P/\mathrm{bar}})~,
    	\end{equation}
        with the pressure $P$ measured in bars and $\dot{v}$ setting the peak amplitude. This forcing is meant to capture the effects of angular momentum transfer to the equator from higher latitudes, for example by Rossby and Kelvin waves \citep{Showman2011}.
    	
        Equation (\ref{eq:induction}) is the $x$-component of the induction equation with $\eta$ the temperature-dependent MD,
    	%
    	\begin{equation}\label{eq:diffusivity}
    	\eta(T) = 0.023 \frac{\sqrt{T}}{\chi_e} {\rm m^2~s^{-1}},
    	\end{equation}
    	taken from \citet{Menou2012a} and based on the results of \citet{Draine1983}. The ionization fraction $\chi_e$ is obtained from the Saha equation \citep{Rogers2014b} adopting solar abundances as given in \citet{Lodders2010}.
    	Equation (\ref{eq:temperature}) is the heat transport equation with specific heat capacity $c_p$ and thermal conductivity
        \begin{equation}\label{eq:thermal_conductivity}
    	    {\bar \chi} = \frac{16 \sigma T^3}{3\kappa_{\rm th}\rho},
    	\end{equation}
        where $\sigma$ is the Stefan-Boltzmann constant.
        The irradiation flux is 
        \begin{equation}\label{eq:flux_irradiation}
            F_{\rm irr} = F_s \exp(-\kappa_v P/g)~,
    	\end{equation}
    	where $F_s$ is the flux from the host star incident on the atmosphere.  
        In equations (\ref{eq:thermal_conductivity}) and (\ref{eq:flux_irradiation}), $\kappa_{\rm th}$ is the Rosseland mean opacity (assumed constant), dominated by opacity at infrared-wavelengths, where most of the heat from the planet is emitted, while $\kappa_v$ is the visible opacity, where the spectral flux of the host star peaks. The thermal conductivity is used to set the dynamic viscosity via an assumed Prandtl number $\mathrm{Pr}$:
    	\begin{equation}
    	    {\bar \mu} = \frac{\Pr \bar \chi}{c_p}.
    	\end{equation}
        The last two terms on the RHS of Equation (\ref{eq:temperature}) are the viscous ($\phi_\nu$) and ohmic ($\phi_\eta$) dissipation functions.

	\subsection{Boundary and initial conditions}\label{sec:BCs&ICs}

    	We impose corotation with the core of the planet at the base, $u_x=0$, and set vanishing mechanical stresses at the top boundary, $\partial u_x/\partial z = 0$.
    	For the magnetic field, we set $\partial B_x/\partial z = 0$ at the lower boundary, so as to ensure vanishing magnetic stress. At the outer boundary, we opt for a vacuum boundary condition, implying $B_x = 0$.
        For the temperature, we emulate a constant flux of energy coming from the core, $F_0$, and a flux coming from the host star, $F_s$, which is absorbed at depths controlled by the visible opacity, $\kappa_v$, as prescribed by Equation (\ref{eq:flux_irradiation}). We therefore set the inner temperature boundary condition at a constant flux corresponding to the sum of the two flux sources. The upper boundary is kept at a constant temperature $T_0$. 
     
        We set the heat flux from the core as $\sigma T_\mathrm{int}^4$ where $T_{\rm int}=150$~K for all models, following M12, and then parametrize the thermal profile with $T_{\rm eq}$ and $\kappa_{\rm th}$.
        For each choice of parameters, we set the values of $F_s$, $T_0$, and $\kappa_v$ by comparing with the temperature profiles presented in M12 which are given as a function of $T_{\rm eq}$. We solve the steady state form of Equation (\ref{eq:temperature}) with only the first three terms, i.e.~we ignore any viscous and ohmic heating, jointly with the hydrostatic balance equation, using a ODE solver from SciPy\footnote{The function is {\tt solve\_ivp} from the {\tt integrate} module. The documentation can be found at {\tt https://docs.scipy.org/doc/scipy/reference/generated/ scipy.integrate.solve\_ivp.html}}, a Python package, using an explicit Runge-Kutta method of order 5. We then determine the values of $T_0$ and $F_s$ as a function of $T_\mathrm{eq}$ by fitting the temperature profiles shown in M12, and adopt a value $\kappa_v=6.93 \times 10^{-4}$ m$^{2}$ kg$^{-1}$ for all models that reproduces the pressures at which the heat is deposited in M12. 

    	The same initial conditions are used throughout: solid body rotation at the lower boundary rate $u_x=0$, and purely radial magnetic field, i.e. $B_z = B_0$ and $B_x = 0$. The initial temperature profile is found jointly with the boundary conditions, as described in the previous paragraph, but we also let the solution from the ODE solver relax on our simulation grid. During this relaxation process, the barred variables in Equations~(\ref{eq:movement})--(\ref{eq:temperature}) are allowed to adjust to the dissipationless steady-state temperature profile, but remain fixed thereafter. The initial MD is then simply given by Equation (\ref{eq:diffusivity}).
    	
	\subsection{Numerical scheme}\label{sec:numerical_methods}

        \begin{figure}[htpb]
    	    \centering
    	    \includegraphics[width=0.9\linewidth]{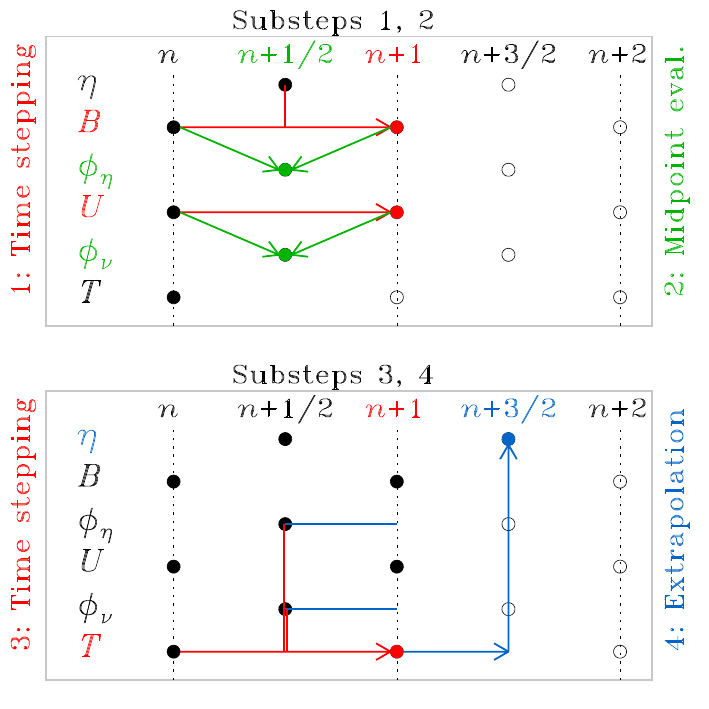}
    	    \caption{Schematic depiction of our 4-substeps time stepping scheme $n\to n+1$ (see text). Solid (open) black circles indicate variable values known (unknown) at the beginning of step $n$, and solid colored dot denote variable values being computed during the substeps, as color-coded. Red (substeps 1 and 3) indicates Crank-Nicolson and/or Euler explicit stepping using known mid-step values of MD and source terms; green (substep 2) indicates a posteriori mid-point evaluation. Blue (substep 4) represents Taylor linear extrapolation from $n+1$ to $n+3/2$, followed by calculation of the MD at $n+3/2$, needed for substep 1 of the subsequent time step.}
    	    \label{fig:extrapolation_scheme}
    	\end{figure}
        
        The numerical solution of Equations~(\ref{eq:movement})--(\ref{eq:temperature}) requires the MD $\eta$ and dissipation functions $\phi_\nu$, $\phi_\eta$ to be advanced in time. Because of the strongly nonlinear dependency of $\eta$ on $T$, and of the dissipation functions $\phi_\nu$ and $\phi_\eta$ on flow and field gradients, simply evaluating these quantities at the current (known) time step $n$ when advancing Equations~(\ref{eq:movement})--(\ref{eq:temperature}) to $n+1$ can lead to runaway numerical instabilities. Consequently, we opted to subdivide each time step into four substeps, as depicted schematically in Figure \ref{fig:extrapolation_scheme}. In substep 1, we first advance the magnetic field and azimuthal velocity from time step $n$ to $n+1$ using a mid-point value for the MD, so as to ensure second-order accuracy in the spirit of centered finite differences. Equations (\ref{eq:movement}) and (\ref{eq:induction}) are advanced using a Crank-Nicolson scheme for the linear terms and Euler explicit for the nonlinear term. In substep 2, we use these newly calculated values of $B$ and $u$ at $n+1$ to do a mid-point ($n+1/2$) evaluation of the viscous and ohmic dissipation functions, namely the last two terms on the RHS of Equation~(\ref{eq:temperature}). Substep 3 then advances $T$ from $n$ to $n+1$ via Crank-Nicolson, with mid-point evaluation of all source terms. The final substep 4 extrapolates $T$ at mid-step, from $n+1$ to $n+3/2$, via a truncated Taylor series, i.e., at every grid point $k$ we write
        \begin{equation}\label{eq:Taylorexp}
            T_k^{n+3/2}=T_k^{n+1}+\frac{\Delta t}{2}\frac{\partial T}{\partial t}\Big|_{k,n+1}~,
        \end{equation}
        the temperature derivative at $n+1$ being computed by evaluating the RHS of Equation~(\ref{eq:temperature}) at $n+1$ for the linear terms, and $n+1/2$ for the dissipation functions. The MD at $n+3/2$ is then computed via Equation~(\ref{eq:diffusivity}). This completes the $n\to n+1$ time step, and the process can now repeat anew to advance from $n+1$ to $n+2$.

        Additionally, due to the boundary conditions imposed on the magnetic field, the ohmic heating is at its maximum at the top boundary. This raises issues, as we are keeping the temperature fixed at this end of the domain. To avoid buildup of a very thin thermal boundary layer, we have introduced a progressive ramp-down to zero of the viscous and ohmic heating terms in the outer 20\% of the grid, following a $\propto -z^3$ dependency so as to maintain continuity of the second derivative in temperature. We have verified that this procedure does not affect significantly the TRI behavior or character of the resulting Alfvénic oscillations.

        We use 100 grid points equally spaced in $z$ for all simulations. This was found to provide an acceptable compromise so as to adequately resolve boundary layers without becoming computationally prohibitive. Our time-stepping algorithm uses a constant time step which allows for suitable time resolution during the burst phase. To achieve this, we have used $\Delta t=10$~s for our simulations. We therefore need a few million time steps to complete a simulation. This represents around 2 core-days on a 4.90 GHz CPU.

	\subsection{Choice of parameter values}
  
        The free parameters for the simulations are the radial magnetic field strength, $B_0$, the velocity forcing amplitude, $\dot{v}$, the thermal opacity, $\kappa_{\rm th}$ and the equilibrium temperature, $T_{\rm eq}$. We opted to keep the Prandtl number small and constant throughout all simulations as ohmic heating is the focus of this study and HJ atmospheres are highly inviscid in our pressure range \citep{Rogers2014b,Beltz2022}. Table \ref{tab:parameters} lists the ranges in different parameter values that were used in our parameter space exploration.

        \begin{table}[htpb]
    		\begin{center}
    			\begin{tabular}{ll}
    				\hline\hline \\[-1em]
    				Parameters & Values \\[0.2em]
    				\hline \\[-1em]
    				$B_0$ (G)             & [10, 100] \\
    				$\dot{v}$ (m s$^{-2}$)   & [0.0001, 0.01]   \\
    				$\kappa_{\rm th}$ (m$^2$ kg$^{-1}$) & [0.0001, 0.001]  \\
                    $T_{\rm eq}$ (K)          & [1000, 1200]    \\
                    Pr                    & [0.01]    \\
    				\hline
    			\end{tabular}
    		\end{center}
    		\caption{\label{tab:parameters} Ranges of the parameters used in the simulations.}
    	\end{table}

        The choice of parameter values is guided by the results of H22. H22 presented a local model with a similar magnetic field and flow geometry to that considered here, that gave estimates for timescales and instability criterion as a function of pressure. Rewritten in dimensional units, the instability criterion that H22 derived is
        \begin{equation}\label{eq:instability_criterion}
    	    \frac{3 \dot{v} \rho^{5/2} \sqrt{\mu_0} \eta^2 E_\eta \kappa_{\rm th}}{16 B_0 T^4} > 1,
    	\end{equation}
    	where $E_\eta$ is the temperature-sensitivity of the MD, as defined in H22. We compute the instability criterion of Equation (\ref{eq:instability_criterion}) across the domain, under the initial conditions, to ascertain in a first approximation which free parameter combinations satisfy Equation (\ref{eq:instability_criterion}) and may result in an unstable system. We find that the H22 instability criterion is a good predictor of instability in our 1D model, and the TRI is robustly present. The fact that radial variations are included slightly changes the region of parameter space conducive to the TRI as compared to the local model results of H22. In addition, the behaviour is more complex since the local model parameters vary significantly across the radial direction, and velocity and magnetic field perturbations interact across a large pressure range.
     


\section{Results}\label{sec:results}


    \subsection{A representative simulation}\label{sec:canonical_sim}

\begin{figure}[htpb]
            \centering
            \includegraphics[width=1.0\linewidth]{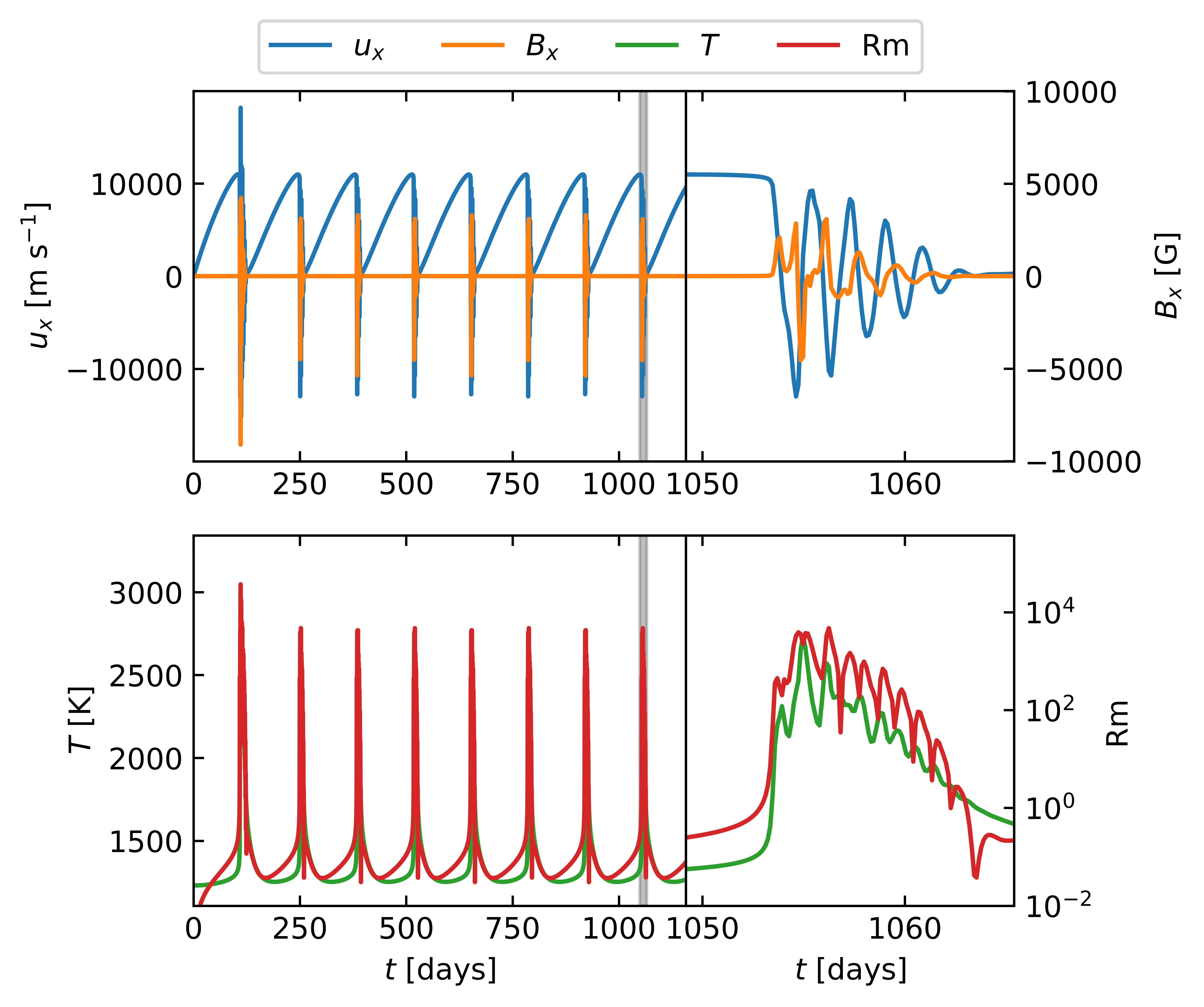}
            \caption{An example of bursting behaviour driven by TRI. We show the time series of velocity, magnetic field, temperature and magnetic Reynolds number in the center of the domain. The model shown has parameters $B_0=30$~G, $\dot{v}=0.004$~m~s$^{-2}$, $\mathrm{Pr}=0.01$, $\kappa_{\rm th}=0.001$~m$^{2}$~kg$^{-1}$ and $T_{\rm eq}=1000$~K. The right panels zoom in on the grey shaded area indicated in the left panels. 
            }
            \label{fig:time_series}
        \end{figure}

        We first examine in detail a representative simulation exhibiting recurrent oscillatory bursts triggered by the TRI. Figure \ref{fig:time_series} shows the time series of velocity, magnetic field perturbation, temperature and magnetic Reynolds number at the center of the domain through a series of seven recurrence cycles of oscillatory bursts (reminiscent of Figure 1 (c) of H22). The local magnetic Reynolds number is defined as $\Rm = u_xH / \eta$, where $H$ is the local pressure scale height.       
        The close-ups on the right hand side of Figure \ref{fig:time_series} focus on a single burst, and highlight the Alfvénic oscillations and their signature $\pi/2$ phase lag between velocity and magnetic field, as well as the exponential relationship between Rm and temperature coming from the ionization fraction, $\chi_e$ (c.f.~Equation~(\ref{eq:diffusivity})). As pointed out by H22, the bursting behaviour is associated with a transition from $\mathrm{Rm}<1$ during the build-up phase between bursts to $\mathrm{Rm}>1$ during the oscillatory phase.

\begin{figure}[htpb]
            \centering
            \includegraphics[width=0.99\linewidth]{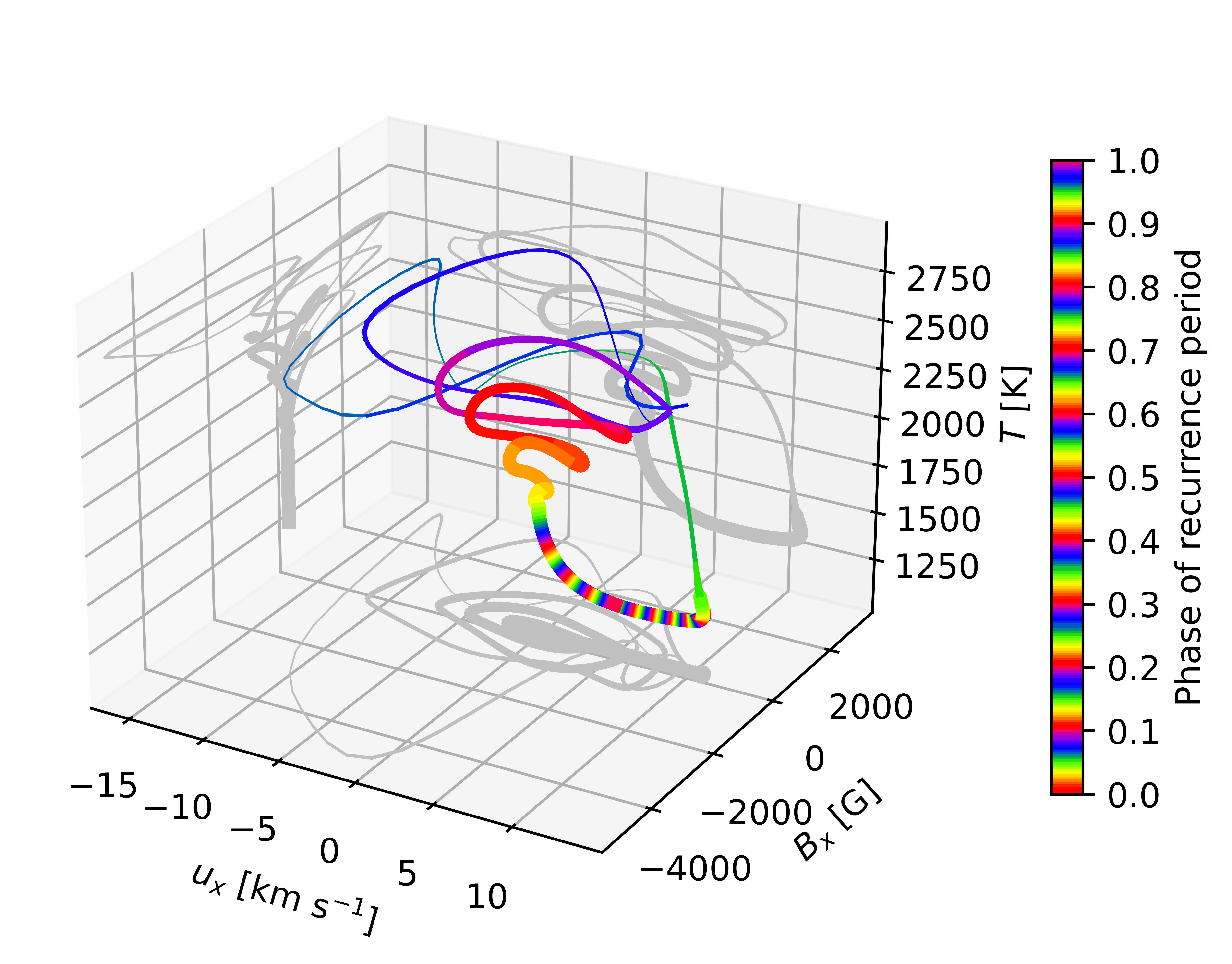}
            \caption{Phase space trajectory at the center of the domain showing one recurrent cycle of the solution shown in Figure \ref{fig:time_series}.
            The thinner the line, the faster the system travels in phase space. The color sequence cycles 10 times through the 134 day recurrence time, so that one run through a color cycle is 13.4 days (about the time required for the system to restabilize after the instability is triggered).}
            \label{fig:phase_space}
        \end{figure}

        Figure \ref{fig:phase_space} shows the phase space $(u_x, B_x, T)$ trajectory of the center point of the domain. The system starts at low velocity and low temperature and moves away from the $u_x=B_x=0$ plane as the velocity increases. The trajectory veers abruptly upwards once the critical temperature for TRI is reached. The initial runaway in temperature is as discussed by M12, but is halted, ultimately, by the dynamical back-reaction of the magnetic field on the flow, similar to the behavior seen in the local model of H22. After the oscillations have damped and the system cooled towards its equilibrium temperature, slow acceleration resumes, building up shear and inducing an azimuthal magnetic field, eventually triggering TRI and another oscillatory burst. A marked separation of timescale exists between the slow buildup phase, typically lasting weeks to months and in which the system spends most of its time, and the very rapid oscillatory burst and decay, spanning a few days.

        \begin{figure}[htpb]
            \centering
            \includegraphics[width=0.99\linewidth]{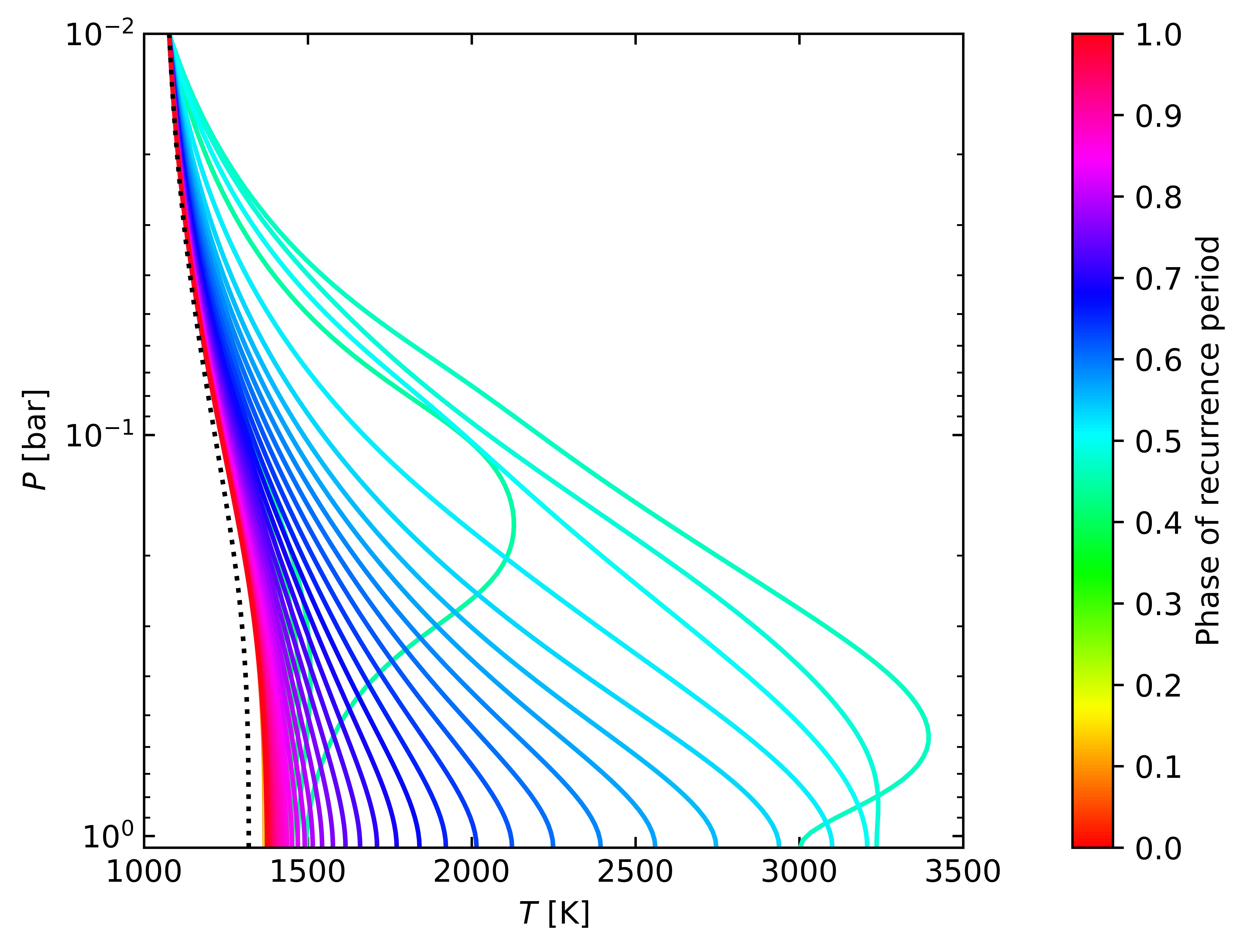}
            \caption{Evolution of the $T$--$P$ profile for the same solution as shown in Figures \ref{fig:time_series} and \ref{fig:phase_space}. The colors represent the phase of the 134 day recurrent cycle.
            The dotted line shows the initial temperature profile. }
            \label{fig:T-P_evolution}
        \end{figure}

        Figure \ref{fig:T-P_evolution} displays the temporal evolution of the $T$--$P$ profile over one recurrence period. The profile changes rapidly once the TRI is triggered. The temperature at the higher pressures can exceed 3000~K at burst peak, as compared to $\sim$1300~K between bursts\footnote{In this example and other cases that reach peak temperatures close to $3000\ \mathrm{K}$, the temperature gradient is steep enough that convection would be expected to occur (not included in our calculation). However, the radial temperature gradient is likely enhanced by the fixed temperature outer boundary condition, and would be reduced with a more realistic outer boundary.}. Such temperatures are expected at these pressures for equilibrium atmospheres with $T_{\rm eq} \approx 2500$~K.

    \subsection{Impact of the diffusivity gradient}

\begin{figure}[htpb]
            \centering
            \includegraphics[width=0.99\linewidth]{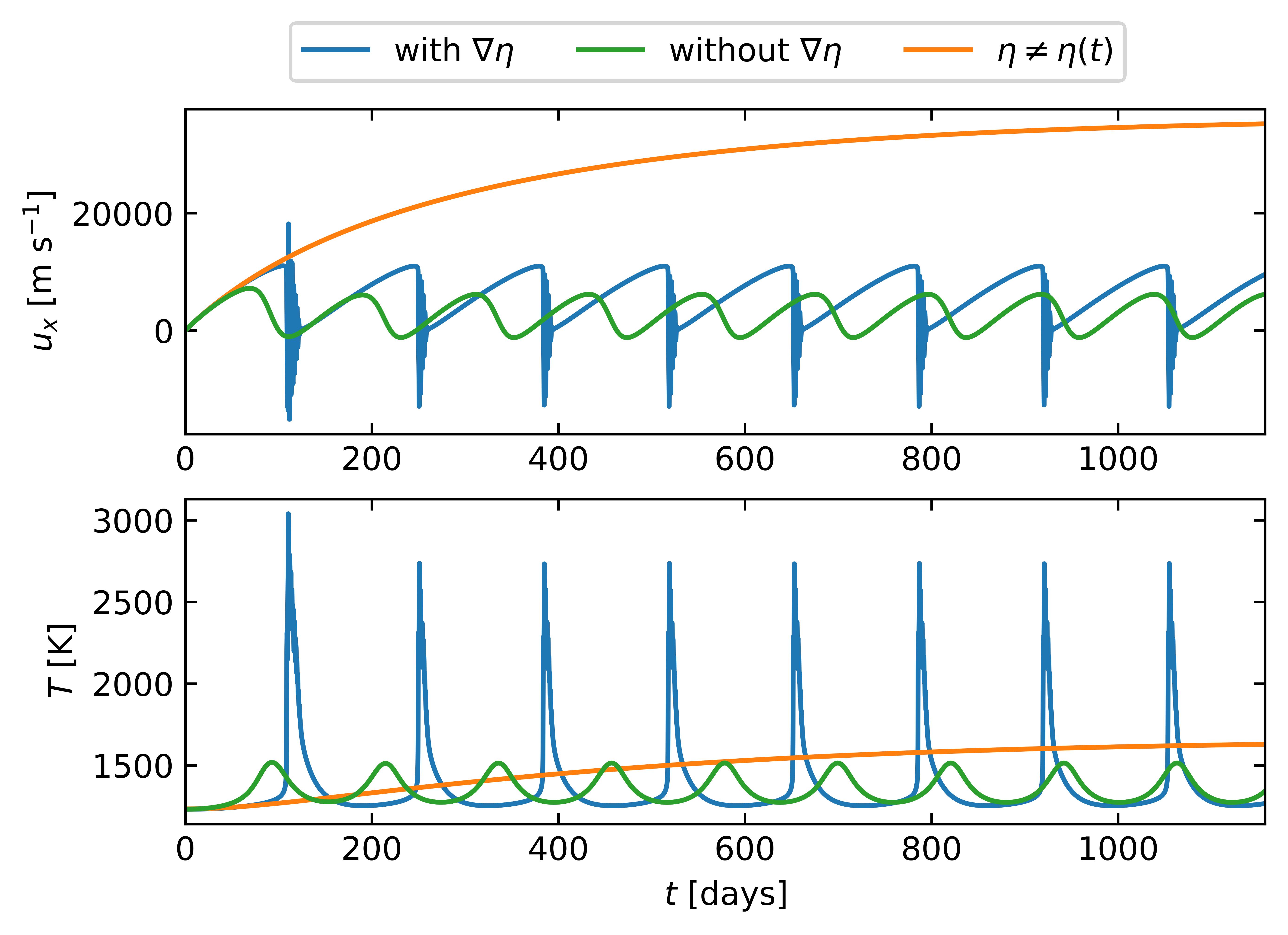}
            \caption{Time series of the mid-point of three simulations with different treatments of the MD. 
            The parameters of the simulation are the same as Figure \ref{fig:time_series}.
            }
            \label{fig:grad_eta_impact}
        \end{figure}

        We find that the (time-evolving) gradient of MD strongly influences the character of bursts and oscillations developing in the simulations. Figure \ref{fig:grad_eta_impact} shows time series for three simulations differing only in their treatment of the diffusivity. In blue, we show a simulation based on the equations as presented in \S \ref{sec:setup}. In green, we show a simulation without the spatial gradient of the MD, i.e. the second term on the right hand side of Equation (\ref{eq:induction}) has been artificially set to zero. In orange, we show a simulation in which the MD is time-independent, fixed to its initial profile.

\begin{figure*}
            \centering
            \includegraphics[width=0.9\linewidth]{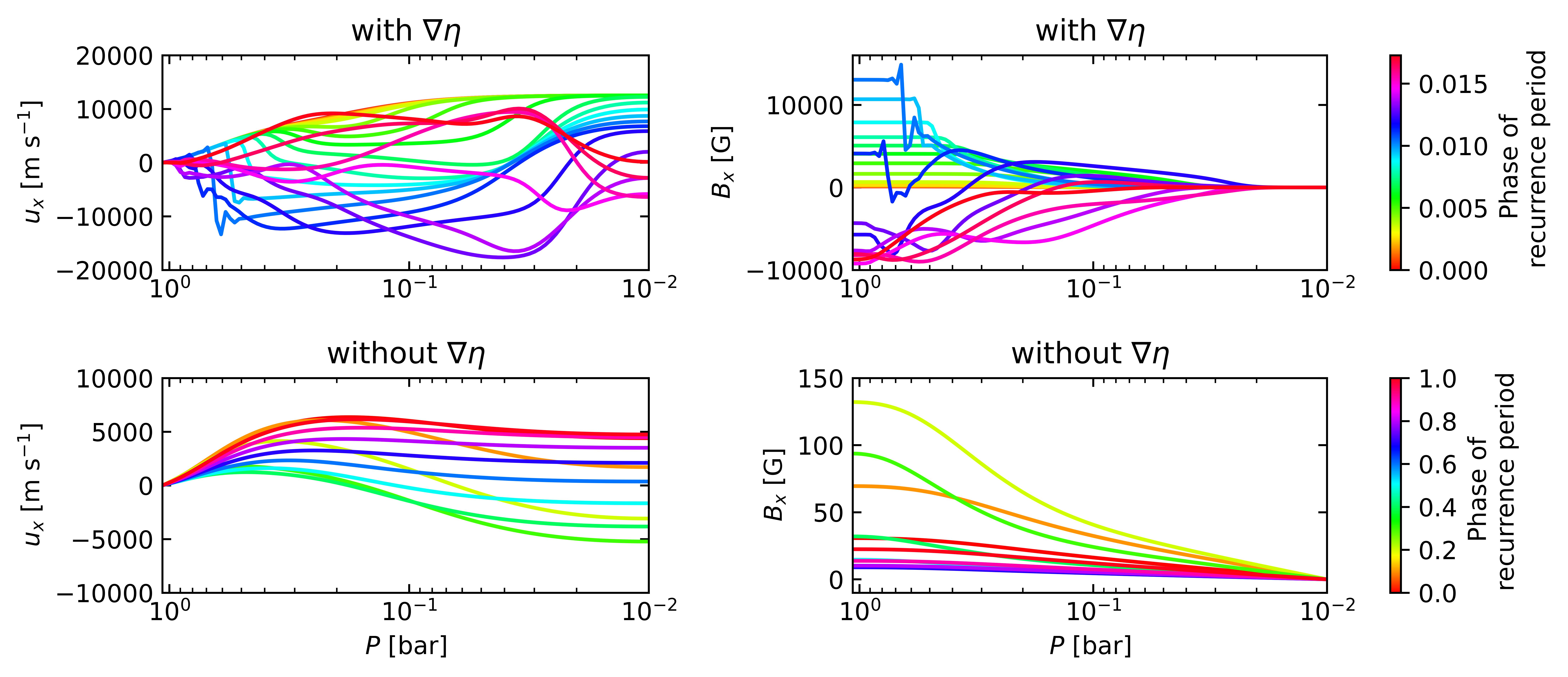}
                \caption{Evolution of the velocity and magnetic field profiles with and without the $\nabla\eta$ term included in Equation~(\ref{eq:induction}). 
                The top plots cover one Alfvénic oscillation (approximately 2\% of the recurrence time) to better show the strong radial gradients during the burst phase; the bottom plots show the full recurrence period.
                The parameters of the simulation are the same as Figure \ref{fig:time_series}.
                We note that during the oscillation with $\nabla \eta$, the simulation is unable to resolve the near-discontinuous front for a brief moment, before it dissipates.}
            \label{fig:grad_eta_impact_profiles}
        \end{figure*}

        While all three simulations show the same initial growth in velocity, the fixed diffusivity simulation (orange) simply grows to a steady-state, as in the local model of H22 and akin to most previous works treating magnetic effects as a drag force. The other two simulations, with time-varying MD, both develop oscillations, but bursting behavior only occurs if the (time-evolving) spatial MD gradient is retained. This can be traced to large spatial gradients of magnetic fields building up
        in the domain, enhancing dissipation and leading to very rapid rise of the temperature (the ``burst''), subsequently transitioning to decaying Alfvén waves. These large magnetic gradients, building up in the lower part of the domain, can be seen in Figure \ref{fig:grad_eta_impact_profiles} (top row), where time-evolving profiles of the velocity and magnetic field are shown during a single post-burst Alfvénic oscillation. Such strong gradients are absent in the simulation with the $\nabla\eta$ term artificially zeroed out (bottom row). The velocity and magnetic field also reaches much smaller values when the $\nabla\eta$ term is removed. For example, at the center of the domain the peak magnetic field is $\approx 5\ \mathrm{kG}$ for the full simulation (Figure \ref{fig:time_series}), but drops to $50\ {\rm G}$ when the $\nabla\eta$ term is removed, and $4\ {\rm G}$ for the fixed $\eta$ case.

        With our geometrical setup and simplifications (i.e., ${\bf B}\equiv B_x(z){\hat x}+B_0{\hat z}$ and $\nabla\eta\equiv\,$d$\eta/$d$z$), the gradient term $-\nabla \eta \times (\nabla \times {\bf B})$ associated with the ohmic dissipation term in the $x$-component of the induction equation becomes equivalent to $(\nabla \eta \cdot \nabla)B_x$, which is structurally akin to advection of $B_x$ by a ``flow'' $\nabla\eta$. This is sometimes referred to as diamagnetic pumping. In our simulations the field is advected inwards, as $\nabla B_x$ is negative during build-up, while $\nabla \eta$ is positive. When the Alfv\'en oscillation starts, large gradients build up near the inner boundary, which lead to the very sharp increase in temperature due to a magnetic field varying very rapidly with depth, producing large amount of ohmic heating.

    \subsection{Region of parameter space exhibiting sustained oscillations and bursts}

\begin{figure}[htpb]
            \centering
            \includegraphics[width=0.99\linewidth]{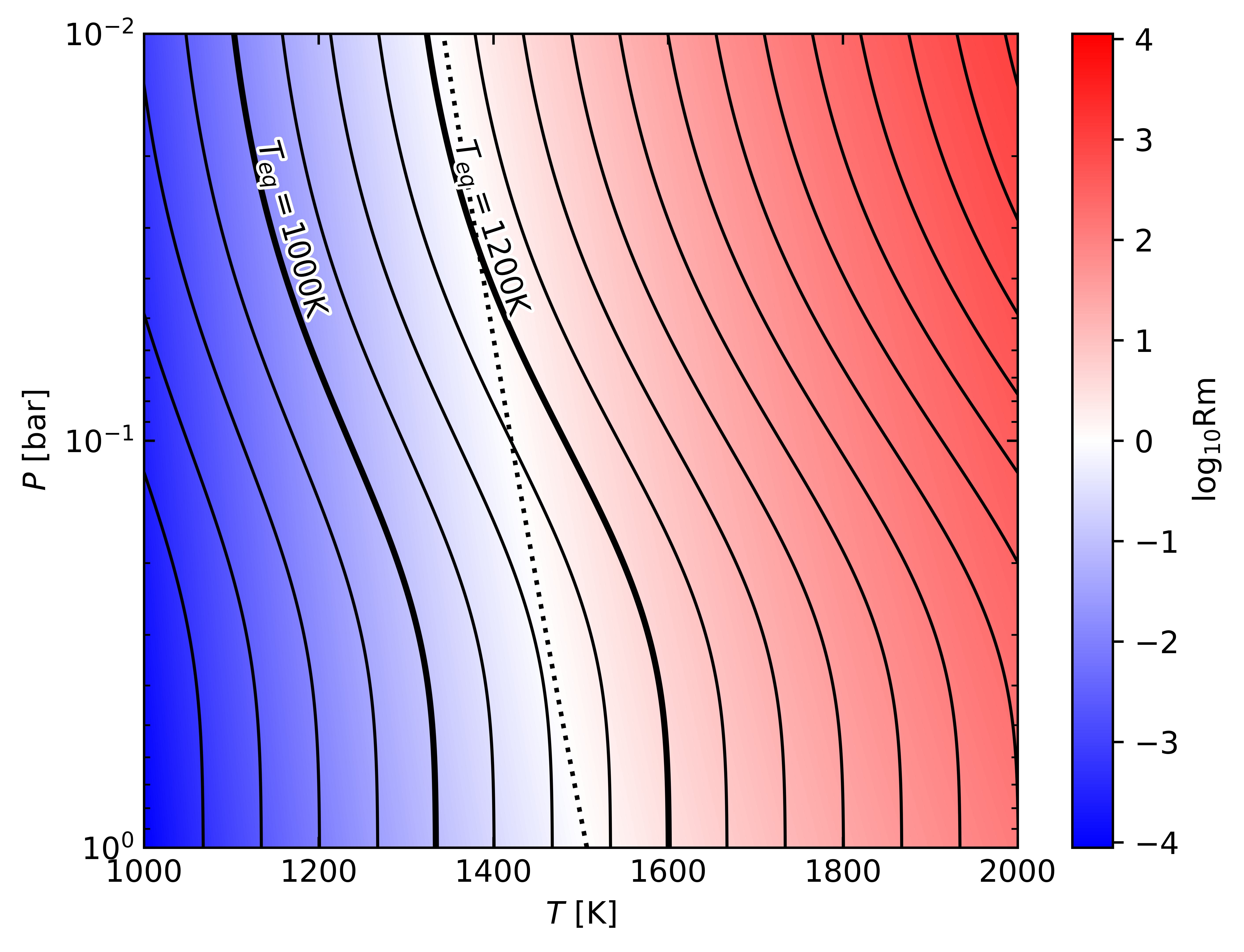}
            \caption{Magnetic Reynolds number $\Rm=u_0 H/\eta$ as a function of pressure and temperature, where we take the characteristic velocity $u_0=5$~km~s$^{-1}$, $H$ is the pressure scale height, and $\eta$ is evaluated at the local temperature and pressure. The solid lines represent typical $T$--$P$ profiles of hot jupiters used in our simulations, with intervals of $T_{\rm eq}=50$~K. The profiles with $T_{\rm eq}=1000$~K and $1200$~K are identified with a label above them. The dotted line shows $\Rm=1$. }     
            \label{fig:expected_RM}
        \end{figure}

        We find that certain choices of parameters are needed for sustained oscillations/bursts. For example, they occur only for equilibrium temperatures $T_\mathrm{eq}\lesssim 1200\ {\rm K}$. This can be understood in terms of the magnetic Reynolds number, since a key characteristic of sustained oscillations is that the system needs to move between weak and strong flux freezing, i.e. between $\Rm<1$ and $\Rm>1$ (H22). Figure \ref{fig:expected_RM} shows an estimate of the magnetic Reynolds number in the $T$--$P$ plane (for this estimate we assume a characteristic velocity of 5~km~s$^{-1}$, a rough average of the velocities in our oscillating solutions). Figure \ref{fig:expected_RM} suggests that the maximum equilibrium temperature allowing TRI-driven oscillatory bursts is around 1200~K, since hotter models have $\mathrm{Rm}>1$ throughout the atmosphere (in agreement with \citet{Dietrich2022}, who also find Rm$=1$ at around $1300$~K). Any system above this equilibrium temperature will be characterized by strong flux freezing from the beginning and will never be able to oscillate around $\Rm=1$. Conversely, colder atmospheres below around 1000~K may never reach $\Rm>1$ as they do not experience enough heating, at least under reasonable velocity forcing.

\begin{figure}[htpb]
            \centering
            \includegraphics[width=0.99\linewidth]{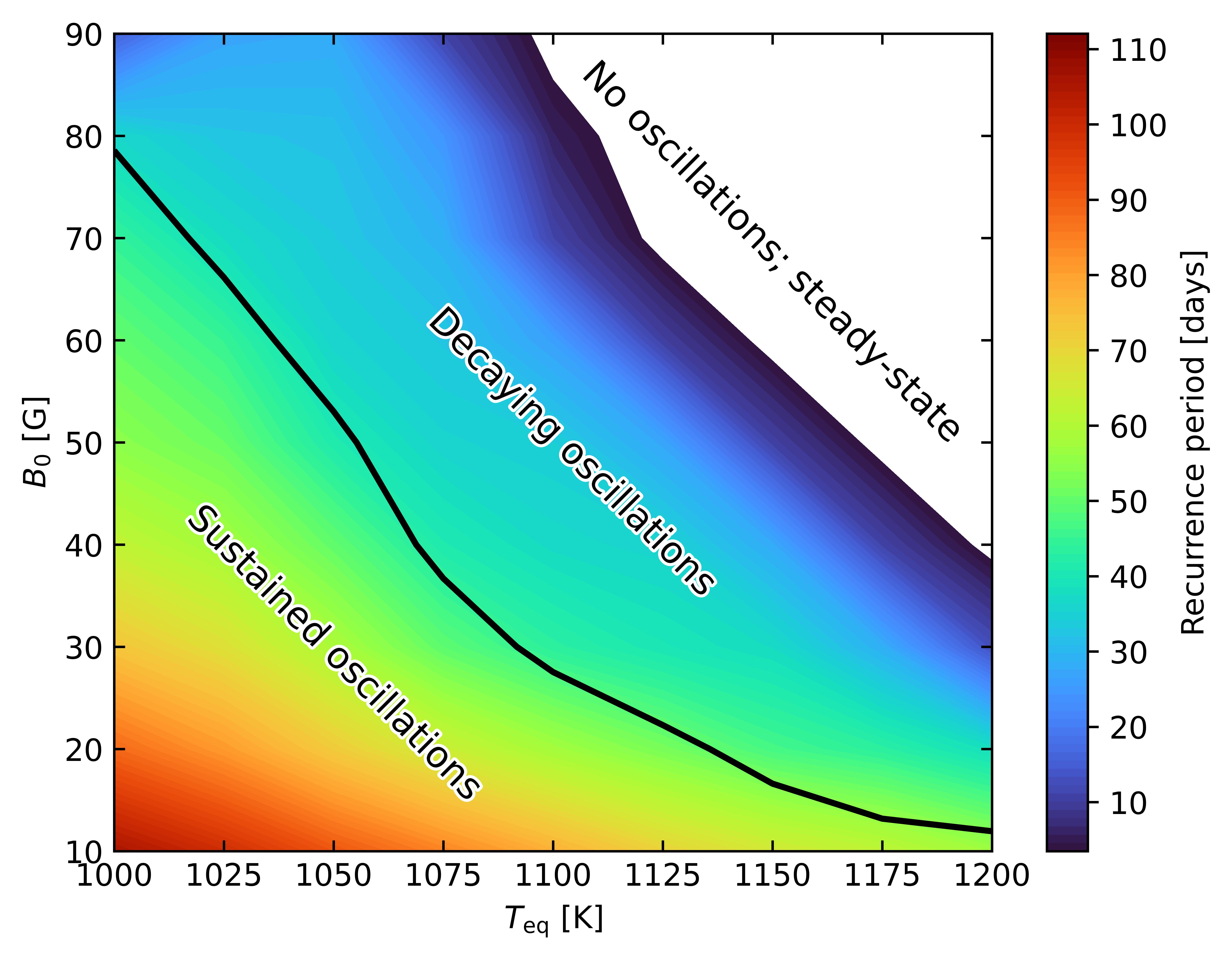}
            \caption{Recurrence period in the instability region for systems with $\Dot{v}=0.01$~m~s$^{-1}$ and $\kappa_{\rm th} = 0.0005$~m$^{2}$~kg$^{-1}$ and different values of equilibrium temperature and radial background magnetic field. Three different regions have been identified corresponding to the behavior of the systems.}
            \label{fig:instability_region}
        \end{figure}

        Figure \ref{fig:instability_region} shows the different regimes in the $T_\mathrm{eq}$--$B$ plane. We plot the recurrence period of the TRI, and take $\Dot{v}=0.01$~m~s$^{-1}$ and $\kappa_{\rm th} = 0.0005$~m$^{2}$~kg$^{-1}$. Instability and resulting oscillations are favored by weaker radial magnetic fields and lower equilibrium temperatures. The cooler temperatures are expected from Figure \ref{fig:expected_RM}. Weaker radial fields require less forcing to be stretched azimuthally against the opposing Lorentz force, which is necessary to generate electrical currents and ohmic heating, the main heating contributor. As expected from H22, we have identified two different regimes in this figure: sustained oscillations, and systems that reach steady-state either by decaying Alfvénic oscillations or directly reach steady-state. We can see the recurrence period of the oscillations steadily shrinking as the systems approach the larger field and temperature values, even across the black line delimiting the two regimes.

    \subsection{Effect of thermal opacity variations}\label{sec:effect_thermal_opacity}

\begin{figure}[htpb]
            \centering
            \includegraphics[width=0.99\linewidth]{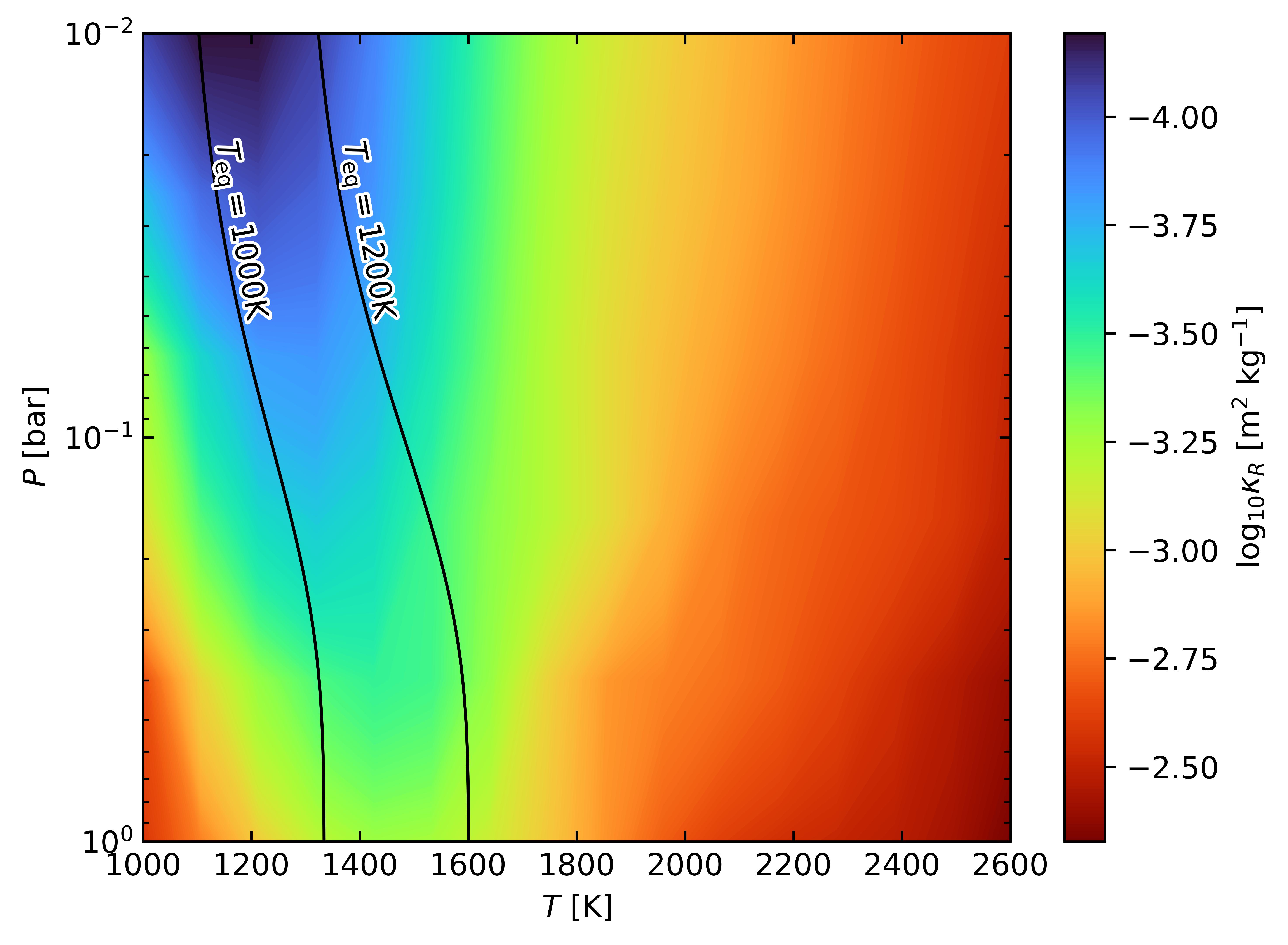}
            \caption{Rosseland mean opacity from \citet{Freedman2008} at solar metallicity in the $T$--$P$ plane. The black lines represent standard $T$--$P$ profiles used in our simulations with $T_{\rm eq}=1000$~K and $1200$~K.}
            \label{fig:Freedman_opacity}
        \end{figure}

        In the simulations presented so far, we assumed for simplicity that the opacity $\kappa_\mathrm{th}$ has a constant value.  In fact, the opacity varies strongly throughout the atmosphere, as can be seen in Figure \ref{fig:Freedman_opacity} where we plot the opacity in the $T$--$P$ plane using the \citet{Freedman2008} opacity tables for dust free gas at solar metallicity. Comparing with the typical atmospheric temperature profiles shown in the Figure, we see that the opacity varies by an order of magnitude within our pressure range, and would also be changing locally during a burst as well as temperature changes.

        To check the effect of opacity variations, we ran a simulation with a depth-dependent opacity profile, but still time independent, using the initial temperature profile to calculate $\kappa_\mathrm{th}$ as a function of depth. We used the following parameters: $B_0=30$~G, $T_{\rm eq}=1000$~K, $\Dot{v}=0.01$, $\mathrm{Pr}=0.01$. As a comparison baseline, we also ran a simulation with a constant thermal opacity of $\kappa_{\rm th}=0.0005$~m$^{2}$~kg$^{-1}$ ($\log_{10} \kappa_{\rm th}=-3.3$), which corresponds to the opacity at around 0.5~bar for this equilibrium temperature. We find that the overall effect of the depth-dependent opacity is that the outer part of the envelope is not as insulating as with a constant opacity. Therefore, higher velocities are needed to reach the critical temperature, thus increasing the recurrence time of the TRI. Otherwise, the two simulations are qualitatively very similar.
         
        HJs may have larger metallicities than solar \citep{Welbanks2019}. \citet{Freedman2008} also presents the values of the mean Rosseland opacity for a metallicity of $[M/H]=0.3$. On average, for temperatures above 1000~K, the opacity with enhanced metallicity is larger by a factor of 1.7. Furthermore, the opacity presented by \citet{Freedman2008} are for dust and cloud free gas. However, the temperature range at which the TRI operates is in a regime where both dust and clouds are expected, which would enhance the opacity of the atmosphere. The opacity presented in Figure \ref{fig:Freedman_opacity} is to be considered as a lower limit of the expected opacity in the atmosphere of a HJ. 

        Moreover, the opacity enters the calculation of the thermal conductivity $\bar{\chi}$ (see Equation~\ref{eq:thermal_conductivity}), which already varies with depth by a few orders of magnitude via its dependence on density and temperature. Introducing a depth dependence on opacity only adds another order of magnitude in the range of thermal conductivity ($\bar{\chi} \sim 1/\kappa_{\rm th} \bar{\rho}$; Equation~\ref{eq:thermal_conductivity}), and both $\kappa_{\rm th}$ and $\bar{\rho}$ go down as pressure goes down). While not an insignificant variation, considering the uncertainties regarding this quantity in HJs atmospheres, the use of a constant thermal opacity is an acceptable approximation.

\section{Discussion}\label{sec:discussion}

        We have presented a model of a forced azimuthal flow in the equatorial region of a HJ which captures the interplay between the dynamics associated with magnetic torques and thermal evolution due to ohmic heating. We find that self-sustained Alfvén oscillations are excited in the range of equilibrium temperatures $T_\mathrm{eq}\approx 1000$--$1200\ \mathrm{K}$ for magnetic field strengths of tens of Gauss (Figure~\ref{fig:instability_region}). The oscillations occur in bursts, separated by long periods in which the magnetic field slowly winds up before a TRI is triggered (Figure~\ref{fig:time_series}). The subsequent increase in temperature increases the magnetic Reynolds number, the flow couples to the magnetic field, and oscillations occur. The need for the Rm to transition from below 1 to above 1 during the cycle sets the temperature range in which oscillations occur, ie.~the temperature of the atmosphere needs to be such that Rm~$\sim 1$ (Figure~\ref{fig:expected_RM}). 

        We find that the spatial gradients of MD strongly modify the radial profile of the Alfvén oscillations (Figure~\ref{fig:grad_eta_impact}), including steepening of the magnetic field profile as perturbations propagate to higher pressure, enhancing dissipation at depth. Diamagnetic pumping like this has previously been found conducive to dynamo action, given a suitable alignment of $\nabla \eta$ and $\nabla \times {\bf B}$ \citep{Busse1992,Petrelis2016,Rogers2017a} (although see also \citealt{Rudiger2022}). Note however that in our model setup, dynamo action is precluded by Cowling's theorem, since both the flow and magnetic field are axisymmetric. In their 3D simulations with a fixed spatially-varying electrical conductivity, \cite{Rogers2017a} found dynamo action for hotter HJs (nightside temperatures $\gtrsim 1400\ \mathrm{K}$). These hotter models would be expected to have large Rm (see Figure~\ref{fig:expected_RM}) enabling efficient dynamo action. At the lower $T_\mathrm{eq}$ values that we consider here, there could be recurrent bursts of dynamo activity, triggered by TRI.

        Although we include the radial structure of the atmosphere, our model makes a number of simplifications. By restricting the geometry to the equatorial plane and assuming axisymmetry, we must impose an external forcing ($\dot v$; Equation~\ref{eq:velocity_forcing}) which models the day-night forcing and transport of angular momentum into the equatorial region by the global atmospheric flow. In addition, we have adopted the simplest possible field geometry (radial field) that supports Alfvén oscillations in our restricted geometry. Full global dynamical models that include the temperature-dependence of the MD will be needed to address these assumptions. We also have assumed constant viscosity and opacity in our model, and this could be relaxed in future work. Although the Prandtl number is small in planetary atmospheres, it would be interesting to study the effect of its time-dependence during the oscillation cycles. Including the temperature-dependence of opacity would give a more accurate determination of the properties of the oscillations (\S \ref{sec:effect_thermal_opacity}).

        One impact of including additional physics in the model may be smaller oscillation amplitudes compared to those we find here. For example, we find that the zonal flows reach supersonic velocities during the burst for some choices of parameters. In the evolution shown in Figures \ref{fig:time_series} and \ref{fig:phase_space}, the peak velocity is above Mach 3. Supersonic velocities are not necessarily unreasonable, considering that they are observed in global hydrodynamical simulations \citep{Cooper2005,Dobbs-Dixon2008,Showman2009,Rauscher2010,Lewis2010,Heng2011}, where the maximum velocities can reach $9$--$10~\rm km\, s^{-1}$ at $2.5\ \rm mbar$, and around $4~\rm km\,s^{-1}$ at $220~\rm mbar$ \citep{Cooper2005,Rauscher2010}. \citet{Dobbs-Dixon2008} also note a peak speed of Mach 2 at the terminator in their simulations. However, we note that our model does not allow for processes, in particular the formation of shocks, that could limit supersonic flows.
        Similarly, additional physics could also intervene to prevent strong winding of the magnetic field, e.g.~magnetic instabilities \citep{Dietrich2022, Soriano-Guerrero2023}. For example, \citet{Dietrich2022} estimate horizontal fields of $\approx 4000\ {\rm G}$ (based on a $10$--$20\ {\rm G}$ planetary field) could arise in KELT-9b from Alfvén oscillations, similar to the field strengths we find here (e.g.~see Figure~\ref{fig:time_series}). However, they estimate that both Coriolis forces and Tayler instabilities would reduce this maximum field strength by about an order of magnitude. Finally, an improved treatment of both the outer and inner temperature boundaries is needed to more accurately determine the peak temperatures reached during the oscillations, in particular, allowing heat to propagate inwards to higher pressure and allowing the outer temperature to respond to changes in the outwards heat flux.


        With the approximations underlying our model in mind, it is interesting to explore the possible implications for known hot jupiters. Figure 2 of \cite{Dietrich2022} shows estimated values of Rm as a function of $T_\mathrm{eq}$ for known hot jupiters. Large $T_\mathrm{eq}$ sources are often considered most interesting from the point of view of magnetism, since they have large Rm. \cite{Rogers2017b} found oscillations in wind velocities and hotspot displacement in simulations of the hot exoplanet HAT-P-7b ($T_\mathrm{eq}=2121\ \mathrm{K}$), and inferred a lower bound on the magnetic field strength needed to reproduce observed variability. Our results suggest that HJs with $T_\mathrm{eq}$ in the range $1000$-$1200\ {\rm K}$ are interesting sources to model and to look at for variability driven by TRI. Possible sources in this temperature range are WASP-69 b, WASP-29 b and HAT-P-12 b on the colder side and WASP-39 b, WASP-34 b, and HD 189733 b on the hotter side\footnote{Equilibrium temperatures were obtained from the \href{http://research.iac.es/proyecto/exoatmospheres/about.php}{ExoAtmospheres database}.}.

        Our results suggest a number of possible observable features of TRI-driven oscillations. Typical recurrence times between bursts range from $2$--$4$ months (Figure~\ref{fig:instability_region}), and the periods of the oscillations are $\approx$ days. While the burst of oscillations is ongoing, ohmic heating raises the atmospheric temperature significantly (Figure~\ref{fig:T-P_evolution}), increasing the atmospheric scale height (by about a factor of 2 in the model shown in Figure~\ref{fig:T-P_evolution}). In addition, the oscillations involve periodic variations in velocity in the atmosphere, and could result in periodic displacements of the hot spot around the substellar point. 

        In summary, the results in this paper and in H22 highlight the need to include the temperature-dependence of MD in atmospheric models, and reveal a new regime of sustained oscillations around Rm~$\approx 1$. A set of 3D simulations that include the temperature-dependence of MD spanning a range of temperatures are needed. This would allow exploration of the full range of behavior, from drag at low Rm, self-sustained oscillations near Rm~$\approx 1$, to dynamo action or magnetic instabilities at large Rm, as well as to make predictions for observables that could be used to constrain the role of magnetism in HJ atmospheres. 

        \vspace{\baselineskip}
        This work was supported by the Natural Sciences and Engineering Research Council of Canada (NSERC) Discovery grants RGPIN-2023-03620 and RGPIN-2018-05278.
        R.H., A.C., and P.C. are members of the Centre de Recherche en Astrophysique du Québec (CRAQ).

\bibliography{mybib}{}
\bibliographystyle{aasjournal}



\end{document}